\documentclass[12pt]{iopart}

\usepackage{iopams}

\expandafter\let\csname equation*\endcsname\relax
\expandafter\let\csname endequation*\endcsname\relax

\usepackage{amsmath}
\usepackage{amssymb}
\usepackage{amsfonts}
\usepackage{graphicx}
\usepackage{bm}
\usepackage{subfigure}
\usepackage{float}
\usepackage{color}
\usepackage{cite}
\usepackage{hyperref}
\hypersetup{
	colorlinks = true,
	urlcolor   = blue, 
	linkcolor  = blue, 
	citecolor  = blue
  }

\begin{document}

\title[V Naniyil \textit{et al}]{Observation of collectivity enhanced magnetoassociation of $^6$Li in the quantum degenerate regime}

\author{Vineetha Naniyil\textsuperscript{1}, Yijia Zhou\textsuperscript{1,2,3}, Guy Simmonds\textsuperscript{1}, Nathan Cooper\textsuperscript{1}, Weibin Li\textsuperscript{1,2}, Lucia Hackerm\"uller\textsuperscript{1}}

\address{ \textsuperscript{1}
School of Physics and Astronomy, University of Nottingham, Nottingham NG7 2RD, United Kingdom \\
\textsuperscript{2} 
Centre for the Mathematics and Theoretical Physics of Quantum Non-Equilibrium Systems, University of Nottingham, Nottingham NG7 2RD, United Kingdom  \\
\textsuperscript{3} Graduate School of China Academy of Engineering Physics, Beijing 100193, China }

\begin{abstract}
The association process of Feshbach molecules is well described by a Landau-Zener (LZ) transition above the Fermi temperature, such that two-body physics dominates the dynamics. However, using $^6$Li atoms and the associated Feshbach resonance at $B_r=834.1$ G, we observe an enhancement of the atom-molecule coupling as the fermionic atoms reach degeneracy, demonstrating the importance of many-body coherence not captured by the conventional LZ model. In the experiment, we apply a linear association ramp ranging from adiabatic to non-equilibrium molecule association for various temperatures. We develop a theoretical model that explains the temperature dependence of the atom-molecule coupling. Furthermore, we characterize this dependence experimentally and extract the atom-molecule coupling coefficient as a function of temperature, finding qualitative agreement between our model and experimental results. In addition, we simulate the dynamics of molecular association during a nonlinear field ramp. We find that, in the non-equilibrium regime, molecular association efficiency can be enhanced by sweeping the magnetic field cubically with time. Accurate measurement of the atom-molecule coupling coefficient is important for both theoretical and experimental studies of molecular association and many-body collective dynamics.

\end{abstract}

\maketitle

In the past decades, Feshbach molecules formed via magnetoassociation~\cite{timmermans_feshbach_1999, Hodby05,Abeelen1999, Mies2000, Nature2003} have captured much attention in the study of unitary dynamics~\cite{Bourdel2004}, collective dynamics~\cite{Grimm2004,Pupillo2020}  and many-body effects~\cite{Musolino2019}. Starting from BCS pairs,  deeply bound molecules are created when the magnetic field is tuned across the Feshbach resonance. A simple model that captures the atom-molecule dynamics is a spin-Boson coupled model~\cite{pazy_conversion_2004, pazy_nonlinear_2005, duine_atommolecule_2004}, where BCS pairs and molecules are mapped to spin-half and bosonic particles, respectively. At zero temperature, the spin-Boson model exhibits rich collective, many-body dynamics~\cite{Sun2016, sun_cooperative_2019}.
Combined with established cooling and trapping techniques~\cite{Ketterle1996,Grimm2000}, this opens up opportunities to explore new fundamental physics~\cite{kohler_production_2006, flambaum_enhanced_2007, carr_cold_2009, collaboration_order_2014, hudson_improved_2011}, controlled chemistry~\cite{vkrems_cold_2008, Knoop2010,ospelkaus_quantum-state_2010-1, de_miranda_controlling_2011, ye_collisions_2018} and the quantum simulation of complex many-body systems~\cite{micheli_toolbox_2006, gorshkov_tunable_2011, pollet_supersolid_2010, lechner_classical_2013}.
\begin{figure}[ht!]
	\centering	
	\includegraphics[width=0.6\linewidth]{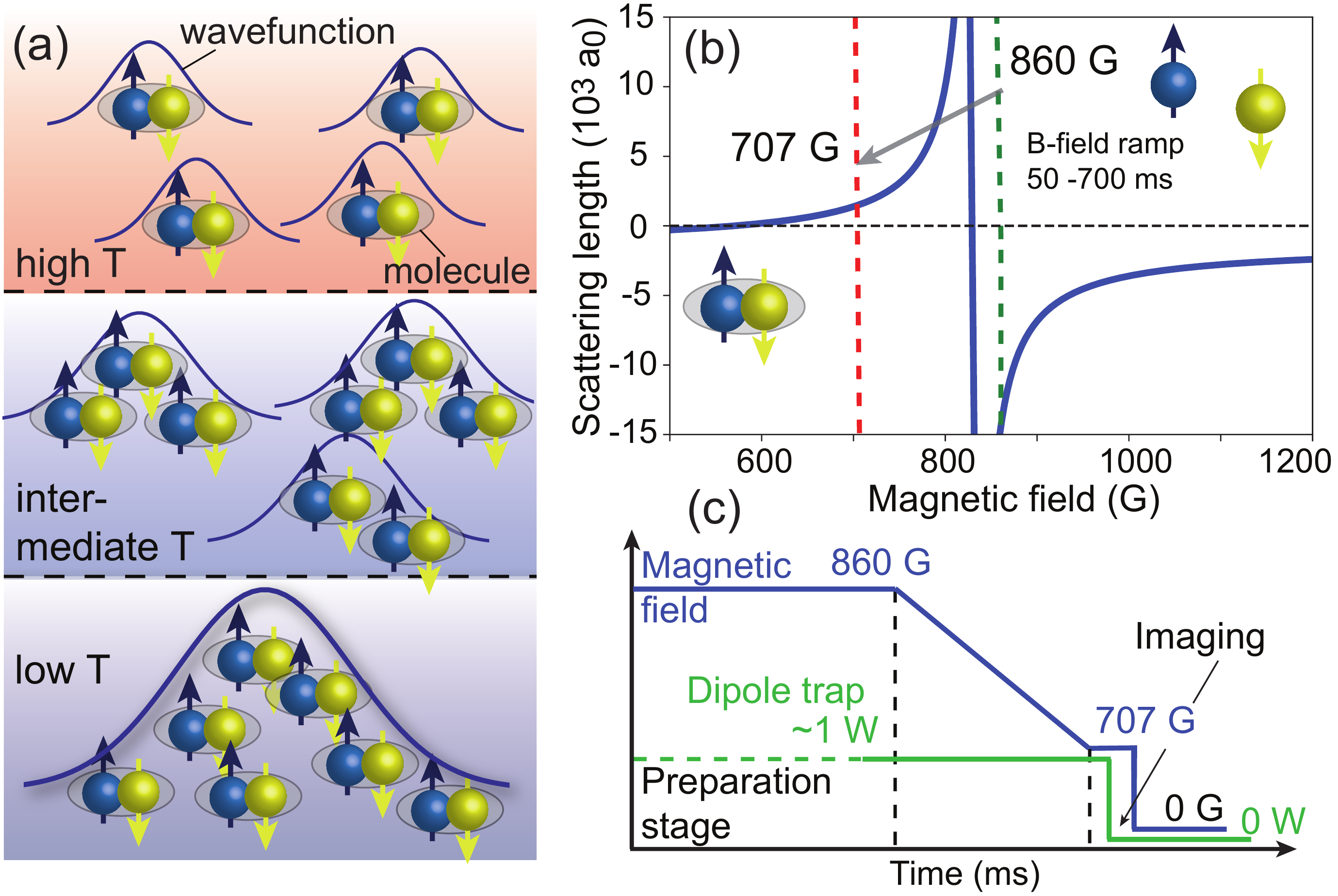}	
	\caption{ \label{fig:diagram} \textbf{Temperature dependence of coherence of Feshbach molecules.} (a) Above the Fermi temperature (top), the thermal wavelength of the atoms is comparable to or smaller than the typical size of the Feshbach molecule. Around the Fermi temperature (middle), the thermal wavelength of atoms is larger than the size of BCS pairs, such that the molecules are partially coherent. At very low temperature (bottom), macroscopic coherence in the atomic gas is established and the $^6$Li molecules form a Bose-Einstein condensate. (b) Broad Feshbach resonance of $^6$Li at magnetic field $B_r=834.1$G, showing the scattering length as a function of the magnetic field for the zero energy collision between an $m_{f}=-1/2$ and an $m_{f}=+1/2$ $^6$Li atom. By ramping down the magnetic field linearly from 860 G to 707 G, pairs of atoms with opposite spin form Li$_2$ molecules. (c) Experimental timing. An ensemble of $^6$Li atoms in $m_f=\pm 1/2$ are prepared with equal populations in a dipole trap. The magnetic field is linearly decreased across the Feshbach resonance to 707 G. We then detect the remaining (unassociated) atoms via absorption imaging. }
\end{figure}

A key parameter by which the dynamics of the spin-Boson model are characterized is the atom-molecule coupling coefficient. The coupling coefficient determines the time scale of the Landau-Zener transition~\cite{goral_adiabatic_2004, Mies2000}, and many-body dynamics~\cite{holland_resonance_2001, Strecker03, pazy_conversion_2004, pazy_nonlinear_2005, szymanska_conventional_2005-1, sun_molecular_2008, altland_nonadiabaticity_2009, yi_atom-molecule_2016} of the atom-molecule system. Understanding the coupling coefficient therefore allows control of the molecular dynamics, including pathways to adiabaticity, and is also crucial for ultracold quantum chemistry. We show here that the temperature dependence of the coupling constant reveals the onset of condensation and provides a smooth connection between the Landau-Zener regime and the fully degenerate regime.

Many theoretical works have shown that the coupling coefficient depends on the magnetic moment of the atom, the background s-wave scattering length and a volume parameter~\cite{timmermans_feshbach_1999, duine_atommolecule_2004}. Here both the magnetic moment and the background s-wave scattering length are constants near the resonance. On the other hand, it has been shown theoretically  that the atom-molecule dynamics becomes collective and should depend on $N$ ($N$ to be the total number of atom pairs), i.e. the association is enhanced by many-body coherence~\cite{pazy_nonlinear_2005, PhysRevA.73.043605, band_adiabatic_2008, sun_molecular_2008, liu_role_2008, Liu2008}. In the fully degenerate regime, the association efficiency is enhanced due to the target state being Bose condensed. This typically requires ultracold temperatures, where the de Broglie wavelength is large, even comparable to the spatial extension of the gas. Experimental and theoretical studies in this regime have provided evidence that the atom-molecule dynamics depends on the entire ensemble. An emerging question is how the coupling coefficient behaves in intermediate temperature regimes $0<T<T_F$ and how it depends on the relevant length scales, such as the de Broglie wavelength and the trap dimensions. A systematic experimental investigation of this dependence has yet to be conducted.

In this work, we investigate collectively enhanced magnetoassociation of $^6$Li atoms below and above the Fermi temperature (figure~\ref{fig:diagram}(a)). The magnetic field is ramped linearly across the broad Feshbach resonance at $B_r=834.1$ G, from BCS pairs ($B>B_r$) to Feshbach molecules ($B<B_r$)~\cite{Salomon2003}, as depicted in figure~\ref{fig:diagram}(b). The fraction of atoms converted to molecules is measured experimentally, as a function of both the temperature of the atomic gas and the sweep rate of the magnetic field. The atom-molecule coupling coefficient is derived from the experimental data through a modified LZ model. We observe that the coupling coefficient increases when the temperature of the atomic gas is lower than the Fermi temperature. An empirical theory based on the mean-field approximation is used to interpret the enhancement of the coupling coefficient as a result of the increased spatial coherence of the molecules, as illustrated in figure~\ref{fig:diagram}(a). With the coupling coefficient, we then theoretically examine the molecular formation dynamics in the quantum degenerate regime. Our numerical simulations based on full quantum mechanics and beyond the mean-field approximation show that molecule conversion can be increased in the diabatic regime when the magnetic field is changed cubically with time. This provides insight into the magnetoassociation process at ultracold temperatures and will be important for the development of quantum technologies based on ultracold molecules.

This paper is constructed as follows. First, we present the experimental protocol and data, which show that the molecular conversion efficiency is temperature dependent. Second, we revisit the LZ model, and make a comparison between the coupling strengths fitted from the experiment and the calculated values based on two-body coupling. Based on the observed Bose enhancement, we provide a theoretical analysis of the enhancement rate, which agrees with the experimental data. Lastly, we propose a non-linear quench scheme, which may further enhance molecular conversion according to our simulation.

\section{Experiment}

In our experiment, we first prepare a cloud of cold $^6$Li atoms in a crossed optical dipole trap. A balanced spin mixture of two hyperfine states is loaded from a magneto-optical trap and evaporatively cooled under a static magnetic field of $B_i = 860.6$\,G, thus placing the atoms on the fermionic side of the Feshbach resonance (see appendix A for details). With the atoms held in the dipole trap, the magnetic field is then ramped linearly across to the bosonic side of the resonance according to $B(t)= B_i - \alpha t$, where $B(0)=B_i$, $B(t_f)=B_f$ and the ramping rate $\alpha = (B_i-B_f)/t_f$. The quench ends with $B_f = 707~{\rm G}$, and $t_f$ is tuned in accordance with $\alpha$. During this process a fraction of the atoms associate into Feshbach molecules. An absorption image of the resulting cloud is taken using light resonant with the D2 line of unassociated atoms of one spin species after a time-of-flight of 1.5\,ms. Due to the molecular binding energy, the imaging light is now detuned by many linewidths (178\,MHz binding energy vs natural linewidth of 6\,MHz) from the corresponding transition in magneto-associated atoms. As a result, the absorption imaging process detects only the unassociated atoms. The molecular conversion efficiency can then be determined by comparing the number of unassociated atoms remaining after the magnetic field ramp to the number present before. For each experimental setting a calibration procedure is applied by ramping back over the resonance, thus dissociating molecules back into atoms (appendix B). A range of different ramping rates $ \alpha$ are employed, such that the total ramping time, $t_f$, varies from $50$ to $700~{\rm ms}$. This whole procedure is then repeated at temperatures between 3.2 $\mu$K and 130 nK, i.e. both above and below $T_\mathrm{F}$. This allows us to explore the molecular association behavior over a broad range of initial temperatures of the atomic gas.

\begin{figure}[htt!]
\centering
\includegraphics[width=0.6\linewidth]{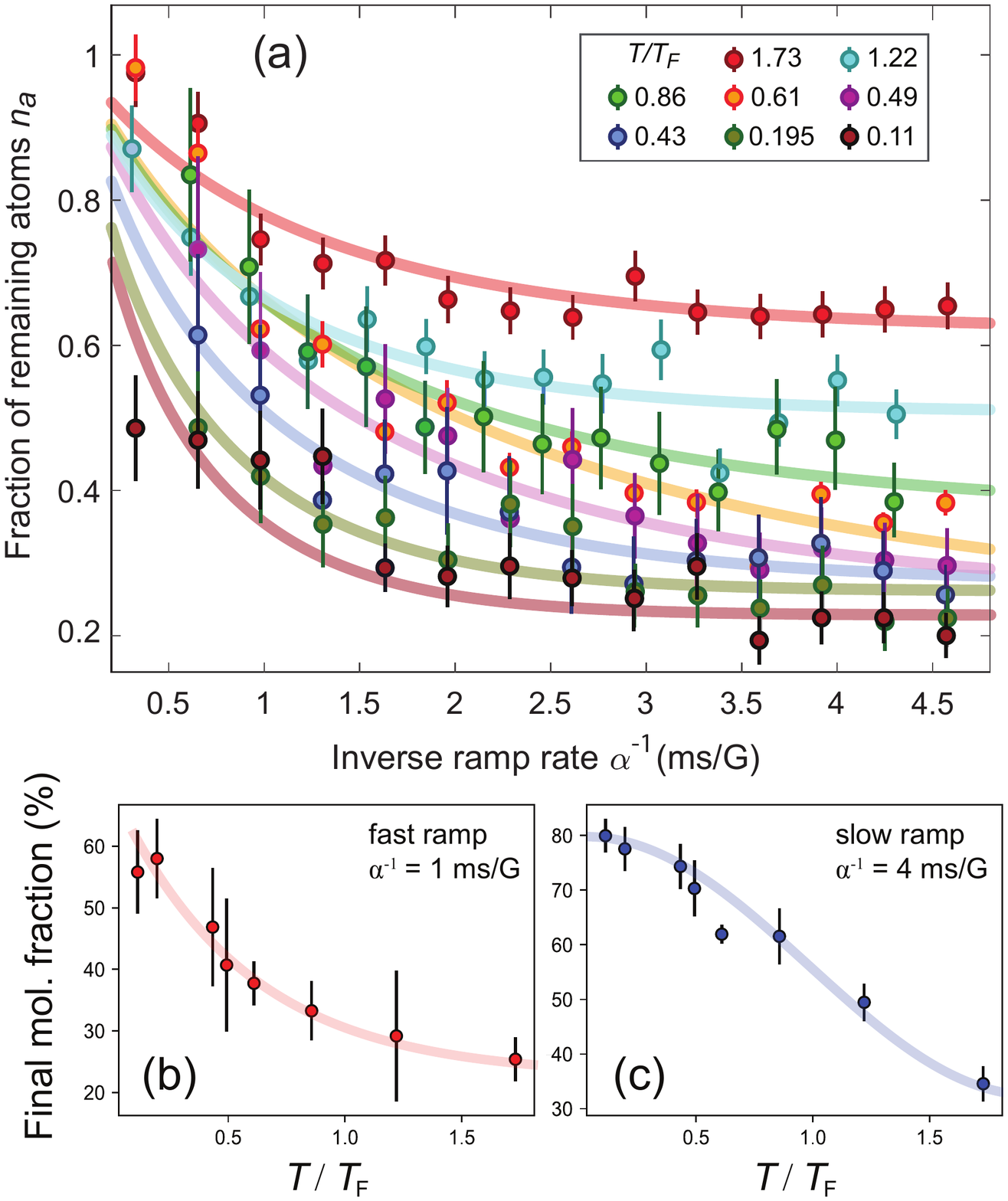}
\caption{(Color online) \textbf{Molecule formation at different temperatures and sweeping rates.} (a) Remnant fraction of non-associated atoms. Different colours refer to different temperatures as given in the legend.  When the inverse ramp rate $\alpha^{-1}$ is low (i.e. fast ramp), the atom fraction is large. Decreasing the ramp speed reduces the fraction of remaining atoms.  The solid lines are fitting results based on equation~(\ref{eq:fiteqn}). The error bars are the standard error of 5 measurements. We show the temperature dependence of the molecule fraction for a fast ramp with $\alpha^{-1}=1$ ms/G  in (b) and a slow ramp with $\alpha^{-1}=4$ ms/G in (c). In both cases, the molecule fraction increases when the temperature is decreased. For the slow ramp, the molecule fraction is around $80\,\%$ when $T/T_F\sim 0.1$. In the experiment, the initial and final magnetic field are $B_i=860.6$\,G and $B_f=707$\,G respectively. }
\label{fig:molecules}
\end{figure}

We first investigate non-equilibrium and equilibrium molecule formation by varying the ramp time.  
The experimental results are shown in figure~\ref{fig:molecules}(a). At a given temperature $T/T_{F}$, the fraction of remaining atoms (as determined via absorption imaging) depends on $\alpha^{-1}$ nonlinearly. A general trend for all temperatures is that the fraction of remaining atoms increases when the magnetic field is changed faster. The fraction of the remnant atoms (molecules) is small (large) when $\alpha$ is small. We find that the remnant atom fraction is non-negligible even in the adiabatic regime. The molecule formation efficiency, i.e. the ratio of the molecules formed to initial atom pairs present, in the adiabatic
limit has been shown to be influenced by multiple collisions~\cite{williams_theory_2006}, and many-body effects~\cite{murthy_high-temperature_2018}. Combined with experimental imperfections, such as spin state imbalance in the initial atom cloud, these are sufficient to explain our observation of sub-unity conversion efficiencies. In the opposite, diabatic regime when $\alpha$ is large, we find the remnant atom fraction increases significantly with respect to $\alpha$ after sweeping the magnetic field.

We find that the molecule conversion rate changes dramatically at different temperatures. In figure~\ref{fig:molecules}(b) and (c), the molecule conversion is shown as a function of the temperature of the atomic gas. When the ramp is fast (figure~\ref{fig:molecules}(b)), the molecule fraction is low at high temperatures and high when the temperature is below the Fermi temperature $T_F=\hbar^2/(2mk_B) (3\pi^2n)^{2/3}$, with $\hbar$ being the reduced Planck constant, $k_B$ the Boltzmann constant, $m$ the mass of a $^6$Li atom and $n$ the total atom number density. The molecule fraction increases monotonically as temperature decreases. Similar dependence is found in the case of a slow ramp (figure~\ref{fig:molecules}(c)). Note that for $T/T_F = 0.61$, this is not the case for very fast ramps. In this regime (fast ramps) the data points for different temperatures are very close to each other, the error bars overlap and the observed variation is attributable to experimental fluctuations. The data points for the two lowest temperatures are also very close to each other as the molecular production efficiency saturates in the low temperature limit (see figure~\ref{fig:molecules}(c)). However, one should note that the overall conversion efficiency is higher in this case. For example, the final molecule fraction (at $T=130$\,nK) increases from less than $60\,\%$ for $\alpha^{-1} = 1\,$ms/G to $>80\,\%$ for $\alpha^{-1} = 4$\,ms/G. The conversion rate is generally high, comparable efficiencies ($>80\,\%$) have been observed and analyzed in other experiments and theories~ \cite{Hodby05, Dobrescu2006, vkrems_cold_2008}. 

\section{Atom-molecule coupling coefficient}

A key parameter to describe the atom-molecule dynamics is the atom-molecule coupling coefficient. While existing theoretical works use different Hamiltonians to describe molecule formation~\cite{javanainen_coherent_1999,timmermans_feshbach_1999, sun_molecular_2008}, the coupling coefficient is generally given by~\cite{PhysRevLett.92.140402,PhysRevA.71.063614}
\begin{equation} \label{eq:g}
	g_c = \hbar\sqrt{ \frac{ 4\pi |a_{bg} \Delta B| \Delta\mu }{ m{\cal V} } }
\end{equation}
where ${\cal V}$, $a_{bg}$, $\Delta B$ and $\Delta\mu$ are the mode volume, background scattering length, resonance width, and the difference in magnetic moments between open and closed channels respectively. In this experiment, $\Delta\mu=2\mu_B$ with $\mu_B$ being the Bohr magnetic moment. At sufficiently low temperatures, the coupling constant is not directly associated with temperature except through ${\cal V}$. To reach different temperatures, the dipole trap frequencies are varied; the mode volume ${\cal V}$ is determined from the trapping frequencies~\cite{sun_molecular_2008}. Equation~\eqref{eq:g} has been widely accepted under low densities or relatively high temperature ($T \geqslant T_F$) conditions~\cite{Hodby05}, where the molecular coupling is truly based on two-body physics. Nevertheless, we note that when $T \gg T_F$, equation~\eqref{eq:g} needs to be modified as the energy (momentum) dependent scattering should be taken into account~\cite{timmermans_feshbach_1999, duine_atommolecule_2004}, but it is beyond our work.

The coupling coefficient $g_c$ is a composite parameter and considered as a constant. It connects the atomic properties ($a_{bg}$, $\Delta \mu$ and $\Delta B$) and external fields (${\cal V}$) of the system. Although it is an important parameter when modeling atom-molecule dynamics~\cite{pazy_conversion_2004, pazy_nonlinear_2005, holland_resonance_2001}, the value of $g_c$ has not been widely discussed and a detailed, temperature-dependent experimental measurement has not been achieved so far. An investigation of the coupling coefficient is important to understand the dynamics of the ensemble under adiabatic vs non-adiabatic timescales, and to study shortcuts to adiabaticity \cite{Torrontegui2013, Guery-Odelin2019, DelCampo2019, Campbell_2020, Koehl_2021}.

\begin{table}[h!]
	\begin{center}
		\caption{ Overview of the calculated and fitted coupling coefficients and the resulting enhancement factor for the respective temperature.} 
		\label{tab:coupling}
		\footnotesize
		\begin{indented}
		\begin{tabular}{@{}ccccc} 
			\br
			\text{Fermi temp.} & \text{Temp.} & \text{Calculated $g$} & \text{Fitted $g$} & Enhancement \\
			$T_F~({\rm \mu K})$ & $T/T_F$ & $g_c~(2\pi\hbar\times{\rm kHz})$ & $g_f~(2\pi\hbar\times{\rm kHz})$ & $g_f^2/g_c^2$ \\
			\mr
			1.23 (1) & 0.11 (1) & 4.79 & 8.78 (42)  & 3.4 (6) \\
			1.30 (1) & 0.20 (1) & 4.62 & 7.68 (14)  & 2.76 (16) \\
			1.39 (1) & 0.43 (2) & 4.48 & 6.46 (22)  & 2.08 (20) \\
			1.48 (1) & 0.50 (2) & 4.30 & 5.31 (29)  & 1.52 (21) \\
			1.53 (3) & 0.62 (1) & 4.30 & 4.51 (42)  & 1.10 (21) \\
			1.61 (3) & 0.86 (2) & 4.19 & 4.55 (27)  & 1.18 (15) \\
			1.69 (2) & 1.23 (7) & 4.69 & 5.95 (29)  & 1.61 (20) \\
			1.86 (3) & 1.73 (4) & 4.73 & 4.64 (47)  & 0.96 (19) \\
			\br
		\end{tabular}
		\end{indented}
	\end{center}
\end{table}

To obtain the coupling coefficient, we note that parameters $a_{bg}$ and $\Delta{B}$ have been measured in a number of experiments~\cite{chin_feshbach_2010}. The mode volume of strongly interacting Fermions in an anisotropic harmonic trap is ${\cal V} = \frac{4\pi}{3} a_{ho}^3 \xi_B^{3/4} \sqrt{24N}$, where the oscillator length is $a_{ho} = (\hbar/m\omega)^{1/2}$ with $\omega$ the geometric mean of the oscillation angular frequency. The parameter $\xi_B$ is the Bertsch factor~\cite{Papenbrock1999, Pitaevskii2016} accounting for the atomic interactions. In the dilute limit, $\xi_B\approx 0.37$ is obtained from Monte Carlo simulations~\cite{Carlson2011}. Using the experimentally obtained total atom number $N$ (and hence ${\cal V}$), we obtain the coupling coefficient $g_c$. The related parameters and the coupling coefficient are summarized in table~\ref{tab:coupling}. The table shows that the coefficient $g_c$ varies only marginally as we change the temperature. Note that the value of $g_c$ depends on the definition of the mode volume ${\cal V}$, while $g_f$ is a parameter that is directly obtained from the experimental data.

The molecule formation can be described by a two-state model~\cite{chin_feshbach_2010}, where two atoms form a molecule through a LZ transition. The two-state model is used to fit the experimental data. The dynamics is governed by the Hamiltonian
\begin{equation} \label{eq:LZ_Hamiltonian}
H = \begin{pmatrix} 0 & g_c \\ g_c & \delta(t) \end{pmatrix},
\end{equation}
where $\delta(t) = \Delta\mu B(t)$. Based on this two-state model, the time-dependent Schr\"odinger equation can be solved analytically. In this idealized scenario,  the molecule conversion rate $\Gamma_m^{\text{LZ}}$ is given  by $\Gamma_m^{\text{LZ}} = 1 - \exp\left[ -2\pi g_{c}^2/(\Delta \mu \alpha)\right]$ in the limit $t\to +\infty$, and the remnant fraction of atoms is $n_a=1-\Gamma_m^{\text{LZ}}$. 

On the other hand, for $T \sim 0$, theoretical investigation of atom-molecule conversion for fully degenerate gases has gone beyond the simple two-state model. It has been found that collective dynamics is expected due to the many-body coherence of the Feshbach molecules. When taking the many-body effect into account, the atom-molecule coupling strength should be scaled by a factor $\sqrt{2-\Gamma_m}$~\cite{Liu2008, liu_role_2008, Li2009}. We adapt this theory to analyze our experimental data in  the low temperature regime. To account for non-participating atoms during the conversion, we introduce a prefactor $\Gamma_{\infty}$ for $\Gamma_m$, which accounts for thermal fluctuations as well as for pairing imbalance and experimental imperfections, such as the inhomogeneity of the laser fields and collisional loss. The molecule conversion rate $\Gamma_m$ is therefore described by 
\begin{equation} \label{eq:fiteqn}
\Gamma_m = \Gamma_{\infty} \left[ 1 - \exp\left( -\frac{2\pi g_f^2}{\Delta\mu\alpha} \mathcal{C}_m \right) \right]
\end{equation}
where $g_f$ is the coupling coefficient obtained by fitting the experimental data (see table~\ref{tab:coupling}), and $\mathcal{C}_m = (2-\Gamma_m)(2-\Gamma_{\infty})$ is a parameter that is determined from a fit to the experimental data. Using \eqref{eq:fiteqn}  and the relation $n_a = 1 - \Gamma_m$, we fit the experimental data from the diabatic to the adiabatic sweeping limit (see figure~\ref{fig:molecules}).  
In the latter case, the adiabatic remnant atom fraction $n_r$ approaches $n_r = 1-\Gamma_{\infty}$ as $\alpha\rightarrow0$, hence $\Gamma_{\infty}$ will depend on the temperature of the atomic gas. This choice reflects the fact that in the experiment, some fraction of the atoms do not participate in the atom-molecule conversion process, due to, e.g., the finite temperature and inhomogeneous density. This fitting equation is consistent with the one studied in Ref.~\cite{Liu2008, liu_role_2008, Li2009}, where the use of the ideal situation (zero temperature and homogeneous density) corresponds to $\Gamma_{\infty}=1$.

By fitting the experimental data with equation~(\ref{eq:fiteqn}), we obtain the fitted coupling coefficient $g_f$ shown in table~\ref{tab:coupling}. In contrast to $g_c$, the fitted coefficient $g_f$ depends on the temperature of the gas. We find that $g_f$ is small at higher temperatures and for $T>T_F$, $g_f$ is nearly identical to $g_c$. At lower temperatures $g_f$ grows gradually and is almost twice $g_c$ when $T/T_F=0.11$. We note that at $T/T_F=1.23$, $g_f$ is slightly larger than its neighboring values. It is unclear what causes this discrepancy. 

\begin{figure}
	\centering
	\includegraphics[width=0.6\linewidth]{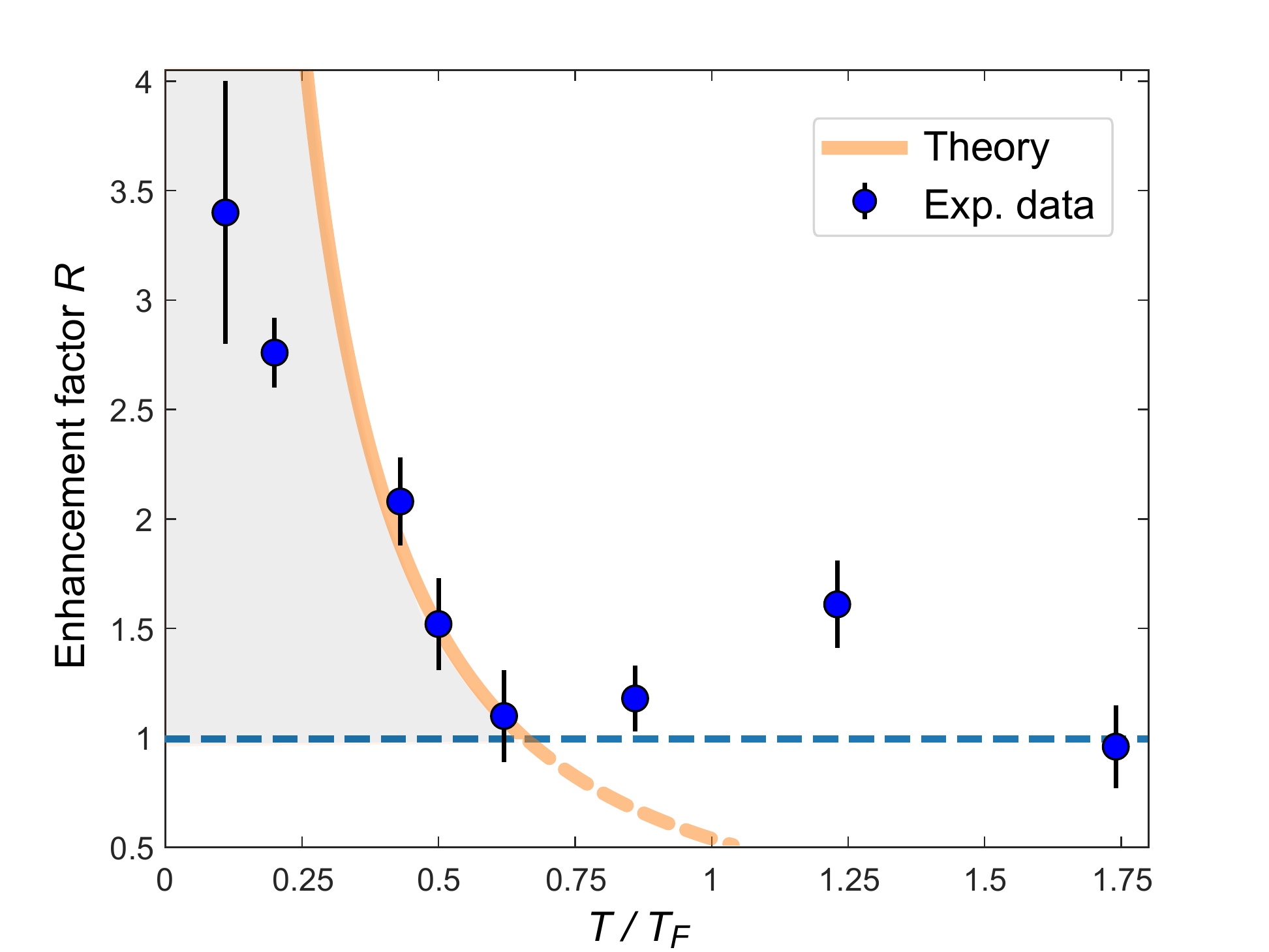}
	\caption{\footnotesize(Color online) \textbf{Temperature dependence of the coupling coefficient}. The orange line represents the factor $R_T$ as defined in equation (\ref{eqn:K}). The theory estimation is bounded by $1$ (blue horizontal line) in the thermal case, and the orange line is dashed where it is lower than this bound.}
	\label{fig:figure3}
\end{figure}

\section{Coherence enhanced molecule conversion}

An interesting question arising from the experimental study is what is the relation between $g_c$ and $g_f$. We first define a geometric factor $R=g_f^2/g_c^2$. Using the experimental data, $R$ is plotted in figure~\ref{fig:figure3} (also see table~\ref{tab:coupling}). It shows that the fitted coupling coefficient becomes large when the temperature is decreased. Note that the system volume and atom number change with temperature. This favours an empirical theory to explain the enhancement of the coupling coefficient. Though the loosely bound Feshbach molecules can spatially extend over the atom cloud, the deeply bound molecules have a much smaller size. When the spatial extension of the molecule is smaller than the thermal wavelength at low temperatures, many-body coherence of the molecules can not be neglected~\cite{javanainen_collective_2004-1, kohler_production_2006}. We can estimate that the coupling coefficient is amplified to be $g_T=\sqrt{N_T}g_{c}$, where $N_T=\rho {\cal V}_T$ is the number of molecules in a thermal volume ${\cal V}_T$ at density $\rho$. Assuming that the molecules have the same temperature as the atoms, their de Broglie wavelength at temperature $T$ is given by $\lambda_T = \hbar\sqrt{2\pi/M k_BT}$, with $M$ being the mass of the Li$_2$ molecule. Thermal volumes of molecules at temperature $T$ are hereafter ${\cal V}_T = \lambda_T^3$. 

For the expected enhancement of the molecule creation, we compare $g_T$ with the coupling coefficient at the Fermi temperature $g_F=\sqrt{N_F}g_c$, where $N_F=\rho {\cal V}_F$, with ${\cal V}_F=3\pi^2\hbar^3\sqrt{(2mk_BT_F)^{-3}}$ corresponding to a Fermi volume at the temperature $T_F$.  As shown by the LZ transition and equation~(\ref{eq:fiteqn}), the molecule fraction depends on $g^2$ when other parameters are given. Inspired by this observation, we therefore characterize the temperature dependence of the molecular production enhancement with a geometric ratio factor,
\begin{equation}
R_T=\frac{g_T^2}{g_F^2} =\frac{2\sqrt{2}}{3\sqrt{\pi}} \left(\frac{T}{T_F}\right)^{-3/2},
\label{eqn:K}
\end{equation}
which leads to a high agreement with the experimental data. 
	
This temperature dependence of the geometric ratio factor shows qualitative agreement with the experimentally fitted $g_f$ when the temperature $T<T_F$, as shown in figure~\ref{fig:figure3}. We thus can interpret the experimental result as follows. 
When the temperature is high $T>T_F$, the thermal volume is smaller than the Fermi volume, ${\cal V}_T<{\cal V}_F$.  The atom-molecule coupling takes place at the two-body level in this high temperature regime. When $T<T_F$, however, the thermal volume is larger than the Fermi volume, which leads to many-body enhanced collective atom-molecule coupling. This means that the molecule conversion efficiency will be higher at lower temperatures, which is consistent with the experimental result (see figure~\ref{fig:figure3}). Note that equation~\eqref{eqn:K} provides a theoretical maximum and experimental values are likely to be below that. Technological restrictions, including the limited lifetime of the dipole trap, three body loss, collisional loss, heating time and the stability of the magnetic fields, are inevitable in low-temperature experiments, which increases the difficulty of reaching the adiabatic regime in our current settings. Moreover, shot-to-shot variations can lower the resulting average value of $g_f$ but not enhance it. Finally, we would like to point out that the model used to interpret the experimental result (i.e. $R$) is empirical. Rigorous theories are thus needed in order to reveal the scaling of the coupling coefficient.

\section{Quantum dynamics of finite systems}

So far we have focused on linear ramps, where the magnetic field is changing linearly across the Feshbach resonance. Here, the maximal molecule conversion rate is only realized in the adiabatic limit. Knowing the coupling coefficient allows the development of alternative schemes to control the molecule association, in particular the achievement of a high conversion rate without the restriction to the adiabatic limit.
Recent studies have shown that shortcuts to adiabaticity can be realized through employing engineered, time-dependent light-atom coupling~\cite{Guery-Odelin2019}. In comparison to the linear association regime, the advantage of a shortcut to adiabaticity is that it provides a fast route to reach the target molecular state while maintaining a high transition probability. Inspired by this advantage, we will explore theoretically the speed-up of the molecule conversion through nonlinear driving, i.e. the magnetic field is swept nonlinearly as a function of time. In particular, we will show that the conversion becomes faster when the magnetic field is changed according to $B(t)\propto t^3$.

At low temperature $T<T_F$, the  Hamiltonian~\cite{javanainen_collective_2004-1,pazy_nonlinear_2005,pazy_conversion_2004} describing the dynamics of molecule formation is given by $H = \sum_j H_j$, where Hamiltonian of the $j$-th pair of atoms reads
\begin{equation} \begin{split}
	\label{eq:Original_Hamiltonian}
H_j &= \delta(t) \hat{b}_j^\dag \hat{b}_j + \sum_k \varepsilon_{jk} \left( \hat{c}_{jk\uparrow}^\dag\hat{c}_{jk\uparrow} + \hat{c}_{jk\downarrow}^\dag\hat{c}_{jk\downarrow} \right)\\
&\phantom{{}=}  + g_c \left( \hat{b}_j^\dag \sum_k \hat{c}_{jk\downarrow}\hat{c}_{jk\uparrow} + \text{H.c.} \right) .
\end{split}
\end{equation}
Here $\hat{b}_j$ ($\hat{b}_j^{\dagger}$) is the bosonic annihilation (creation) operator of a molecule in the $j$-th energy level of the harmonic trap, while $\hat{c}_{jk\sigma}$ ($\hat{c}^{\dagger}_{jk\sigma}$) denotes the annihilation (creation) operator of a fermionic atom with momentum $k$ and spin $\sigma$ ($\sigma = \uparrow,\, \downarrow$). The parameter $\delta(t)=\Delta\mu B(t)$ gives the molecular energy, where $B(t)$ is the magnetic field which is tuned through the resonance point, and $\varepsilon_{jk}$ is the sum of kinetic and potential energy of the atom pair. It is a good approximation to neglect this term when the temperature is low~\cite{pazy_nonlinear_2005}. The atom-molecule coupling coefficient, $g_c$, is given by equation (\ref{eq:g}). At low temperatures, molecules condense into the ground state, and only the harmonic state $j$ with the lowest energy is occupied, i.e. $\hat{b}_j\to \hat{b}$ ($\hat{b}^{\dagger}_j \to \hat{b}^{\dagger}$). To study dynamics of the molecule formation, we propose a general ramping scheme 
\begin{equation}
	\delta(t)/g_{c} = \bar\alpha (g_{c}t)^\nu,
\end{equation}
where $\bar{\alpha}$ is a dimensionless ramping rate and $\nu$ is an odd positive integer (if $\nu$ is even, then $\delta(t)$ becomes non-monotonic). The ramping exhibits power-law dependence on time, and returns to the LZ problem when $\nu=1$. 

\begin{figure}
	\centering
	\includegraphics[width=0.6\linewidth]{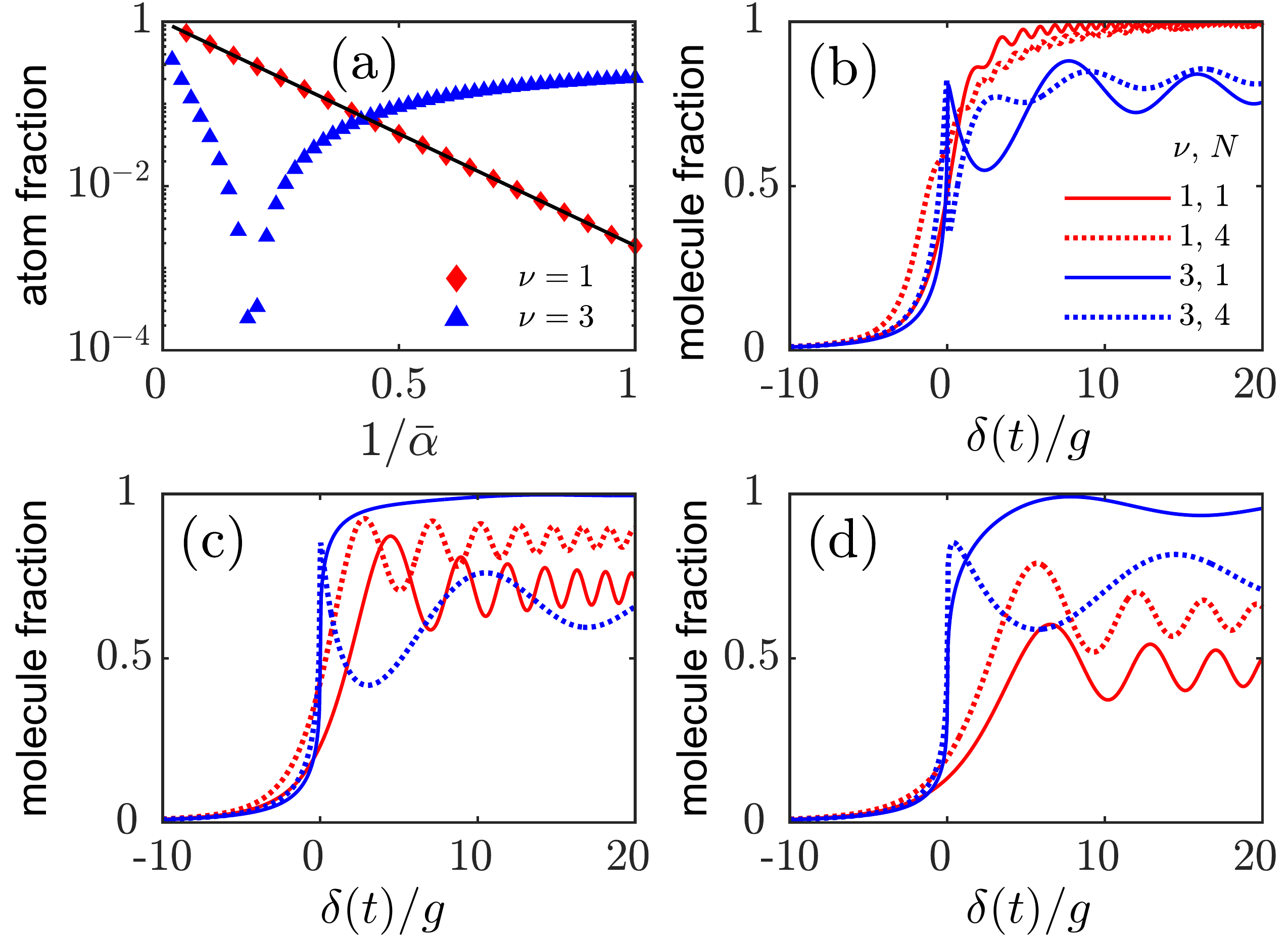}
	\caption{ \footnotesize(Color online) \textbf{Quantum dynamics of the molecule formation}. (a) Comparison between linear (red diamonds) and  polynomial magnetic field ramping (blue triangles) for the remnant atoms at different rate $\bar\alpha$ for a single pair of atoms. The linear ramping follows the Landau-Zener formula (black line). Dynamical evolution of the molecule fraction for (b) $\bar\alpha=1$, (c) $\bar\alpha=5$, (d) $\bar\alpha=10$. When the number of atom pairs is $N=1$, the final conversion efficiency drops dramatically as $\bar\alpha$ increases (b-d,red). However, when $N=4$, the final conversion efficiency is more resistant to the ramping rate (b-d, blue). }
	\label{fig:figure4}
\end{figure}

We simulate the dynamics with the effective two-level model of equation~\eqref{eq:Original_Hamiltonian}, where the system starts with the atomic state and ramps from $\delta/g_c = -50$ to $100$. For a single pair of atoms, the ramping with $\nu=3$ is drastically different from the linear ramping, as shown in figure~\ref{fig:figure4}(a). In case of $\nu=1$, the atom fraction is small only when the ramping is slow. In contrast, the atom fraction in the nonlinear ramping ($\nu=3$) is much smaller than the linear case even when $\bar{\alpha}$ is large (figure~\ref{fig:figure4}(b)-(d)). The minimal (maximal) atom (molecule) fraction appears when $1/\bar\alpha\approx0.2$. The abrupt reduction of the atom fraction happens due to the non-equilibrium dynamics. 

We then simulate the influence of the many-body effects by increasing the number of atom pairs. The results are shown in figure~\ref{fig:figure4}(b-d). For the  linear ramping, the molecule fraction increases gradually with increasing $N$ (see figure~\ref{fig:figure4}(b-d), red). However the conversion efficiency reduces when the ramping becomes faster. 
When $\nu=3$ and $N=4$, the molecule fraction is lower than the $\nu=1$ case in the slow ramping regime (see figure~\ref{fig:figure4}(b)-(d)). In the fast ramping regime (d), the final conversion rate is significantly enhanced when $\nu=3$. The choice of an odd number was made to keep the quench process monotonic, apart from that $\nu=3$ was chosen arbitrarily. This result suggests that  applying the nonlinear ramp in the fast regime is particularly beneficial as it facilitates the molecule formation. 

\section{Conclusion}

Applying a linear magnetic field ramp, we have performed molecule association measurements in the adiabatic and diabatic regime for temperatures between the deeply degenerate and non-degenerate regimes. Our experiment shows that the fraction of atoms associated into molecules increases when both the temperature of the atom gas and the sweeping rate of the magnetic field are decreased. We have measured the atom-molecule coupling coefficient, which increases at lower temperatures and in the adiabatic regime, as a result of many-body coherence. The qualitative trends predicted by an empirical theory agree with our experimental findings, and quantitative agreement appears strong at temperatures slightly below the Fermi temperature. The quantitative differences at even lower temperature indicate that a more sophisticated theory and further experiments are needed. Our study provides a first attempt to accurately measure the atom-molecule coupling coefficient. Exploitation of the coupling coefficient is important to understand the time scale of the molecule association and  might lead to a path for efficient molecule creation through  ramping the magnetic field nonlinearly. For example, our theoretical study shows that cubic ramping enhances the molecule production efficiency even in the diabatic regime. Such nonlinear ramping is thus worth investigating in future experiments. 

\section*{Acknowledgments}	
Daniele Baldolini is thanked for historic contributions to construction of experimental apparatus. This work was supported by the EPSRC Grants EP/R024111/1 and EP/M013294/1 and by the European Commission Grant ErBeStA (No. 800942). WL acknowledges support from the UKIERI-UGC Thematic Partnership (IND/CONT/G/16-17/73), the Royal Society through the International Exchanges Cost Share Award No. IEC$\backslash$NSFC$\backslash$181078, and a RPA grant from the University of Nottingham.

\appendix

\section{Technical details of the experimental procedure}

The generation of the cold atom cloud prior to magnetoassociation begins with a magneto-optical trap (MOT) \cite{Raab1987}. The MOT is loaded via a Zeeman slower \cite{slower_paper}, which slows an atomic beam that is transmitted through a differential pumping stage from a source chamber. Over a 10 s loading cycle, the MOT captures $\sim 2 \times 10^8$ $^6$Li atoms. An additional cooling step, in which the trapping lasers are tuned to half a natural linewidth below resonance (for optimal Doppler cooling), brings the temperature of the atom cloud down to $\sim 300~\mu$K. 

An optical dipole trap is loaded from this cold cloud. A 100\,W fiber laser, operating at 1070\,nm, is used to produce a crossed-beam dipole trap, in which each beam is focused to a waist of 80\,$\mu$m. The crossing angle is 14 degrees. This captures up to $2\times 10^6$ atoms. These atoms are then evaporatively cooled to a regime close to quantum degeneracy to temperatures between $0.1 - 2.0$ T/T$_F$ with total atom numbers between 100000 - 200000 atoms. After the loading,  the dipole trap is first held at constant power for 600\,ms, following which the power in the optical dipole trap is ramped down to the range of tens to hundreds of mW. The end point depends on the final trap depth desired and is reached in a series of linear ramps that collectively approximate an exponential decay of the trapping power. The power is lowered through a combination of reducing the laser current and the use of an acousto-optic modulator. A photodiode is used to measure the optical power that passes through the dipole trap, with servo-controlled feedback to the acousto-optic modulator enabling active stabilization of the dipole trap's depth to its set value. This is necessary to reduce unwanted heating effects arising from small variations in trap depth.

At the end of this evaporative cooling cycle, which lasts $\sim 10$\,s, on the order of 10$^5$ atoms typically remain, at temperatures ranging from tens of nK to several $\mu$K. The cloud is then held at constant trap depth corresponding to trapping frequencies between 622 - 750 Hz (radially) and 74 - 90 Hz (longitudinally).

\begin{figure}
	\centering
	\includegraphics[width=0.6\linewidth]{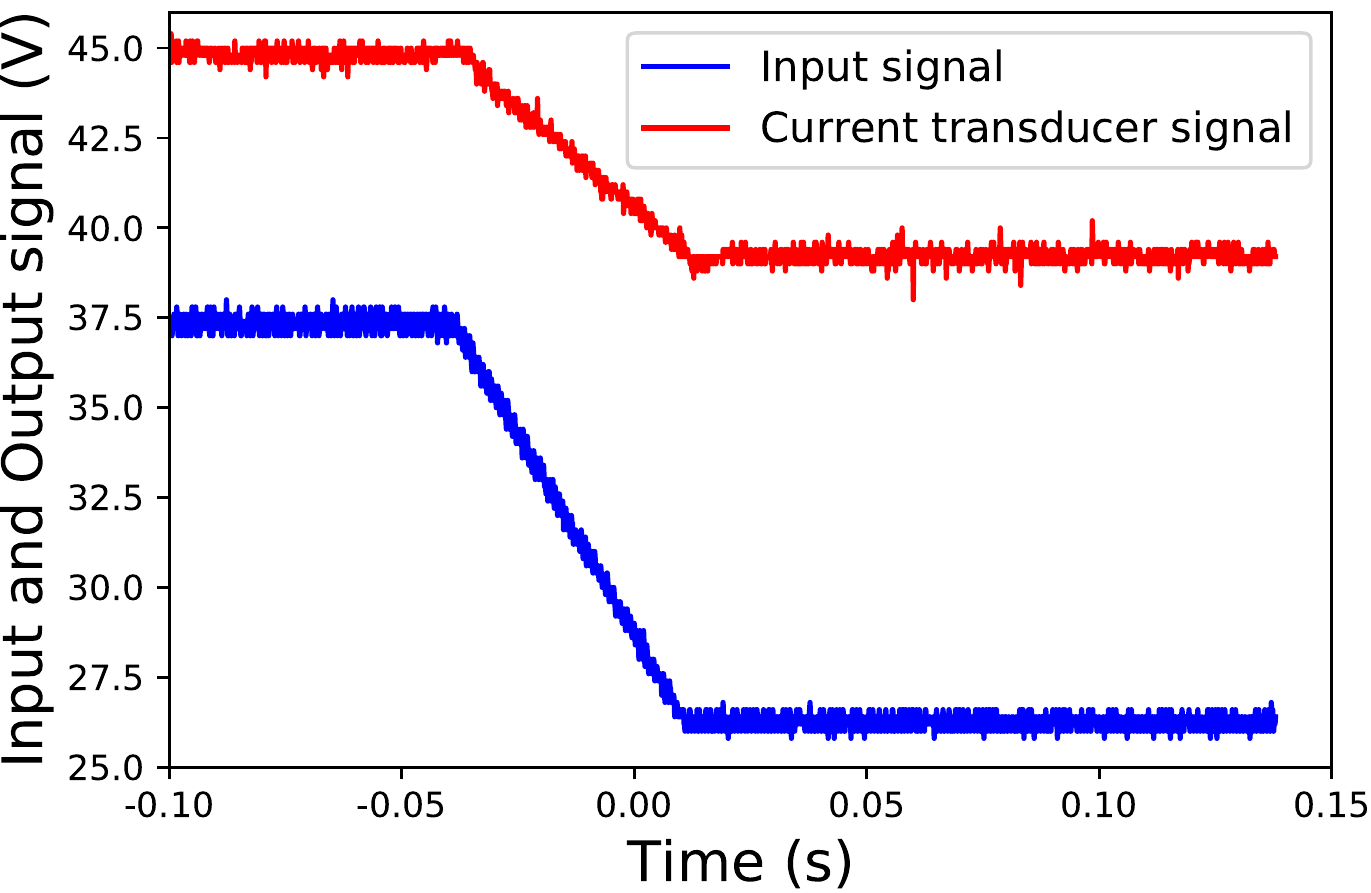}
	\caption{\footnotesize(Color online) \textbf{Magnetic field ramp.} Current transducer signal for a 50\,ms ramp.}
	\label{fig:ramp}
\end{figure}

The magnetic field is then ramped linearly from 860.6 G to 
the BEC side of the Feshbach resonance (707 G). The linear magnetic field ramp is applied through a change in the current in the Feshbach coils as shown exemplary in the figure~\ref{fig:ramp}. 

\section{Determination of molecule fraction via absorption imaging}

To reduce the impact of technical noise sources on the absorption imaging, the atom cloud was released from the dipole trap and allowed to expand for a period of 1 to 2\,ms (depending on exact experimental parameters) prior to imaging. The size of the atom cloud after this period was typically some hundreds of micrometers, which greatly exceeds our imaging resolution of 3 $\mu$m. Each absorption image is background-subtracted and then normalized to an equivalent image taken 50\,ms after the atoms have been dispersed, which greatly reduces the influence of technical noise sources on our data.

We also carry out additional control experiments to account for the effect of loss of unassociated atoms from the dipole trap during the magnetic field ramp. If not properly accounted for, this could cause overestimation of the molecular fraction after the ramp, since we assume that atoms not seen in the absorption image are associated into molecules. We therefore conduct, for each set of experimental conditions under which we take data, a control experiment in which the magnetic field is ramped across the Feshbach resonance and then back again, thus dissociating any molecules that were previously formed. This process is time-symmetric, taking twice as long as the unidirectional ramp, and we therefore assume that the fraction of the atoms remaining after this process is equal to the square of the total fraction remaining (in both associated and unassociated forms) after a unidirectional ramp. This allows us to estimate the reduction in apparent atom number that results from atom loss during the magnetic field ramp under each set of experimental conditions employed. By dividing the apparent unassociated atom fraction that we measure using absorption imaging by this value, we can thus eliminate the systematic bias resulting from atom loss during the magnetic field ramp.

\section{LZ transition of a two-level system}

The molecule formation via sweeping a magnetic field through the Feshbach resonance can be modeled to be a LZ transition. Using a two-state process picture \cite{chin_feshbach_2010,goral_adiabatic_2004}, LZ describes the transition under the Hamiltonian
\begin{equation} \label{aeq:LZ_Hamiltonian}
	H = \begin{pmatrix} \varepsilon & g \\ g & \delta(t) \end{pmatrix},
\end{equation}
where $\delta(t) $ slowly increases from $-\infty$ to $+\infty$ at a constant speed $\dot\delta$. 
Near a Feshbach resonance, $\delta(t) = \Delta\mu B(t)=-\alpha t$, where $\Delta\mu$ is the difference of magnetic moment, and $\alpha$ the sweeping rate. $g$ is the atom-molecule coupling strength. For fermions, it is equal to $g = \hbar\sqrt{4\pi |a_{bg} \Delta_B| \Delta\mu / m} / \sqrt{{\cal V}}$~\cite{duine_atommolecule_2004}. For $^6$Li at $B_r=834.1~\rm{G}$, $\Delta\mu = 2\mu_B$, the resonance width $\Delta_B =-300~\rm{G}$ and the background scattering length $a_{bg}=-1405~\rm{a_0}$. ${\cal V}$ is the mode volume. 

Now we define the wave function $|\psi\rangle=A|a\rangle + B|m\rangle$ where $|a\rangle$ and $|m\rangle$ denote the atomic and molecular state with probability amplitude $A$ and $B$, respectively. The dynamics of $A$ and $B$ are governed by the Schr\"odinger equation,
\begin{eqnarray}
	\dot{A} &=&-ig B, \nonumber\\
	\dot{B}  &=& -igA +i\alpha tB.\nonumber
\end{eqnarray}
To convert the above equations to the standard LZ problem, we make the following transformation, 
\begin{eqnarray}
	A &=& \exp\left[\int_{t_0}^t i\frac{\alpha t}{2}dt\right]a = \exp\left[\frac{i(t^2-t_0^2)}{4}\right]a,\nonumber \\
	B &=& \exp\left[\frac{i(t^2-t_0^2)}{4}\right]b.\nonumber
\end{eqnarray}
The dynamics of $a$ and $b$ is given by
\begin{eqnarray}
	\dot{a} &=& -\frac{i\alpha t}{2} a -i g b,\\
	\dot{b} &=& \frac{i\alpha t}{2} b - i g a.
\end{eqnarray}
We can obtain a second order differential equation of the atomic wave function,
\begin{equation}
	\ddot{a} + \left(g^2-\frac{i\alpha}{2}+\frac{\alpha^2t^2}{4}\right)a=0.
\end{equation}
Using the initial condition $a(-\infty)=0$, we find the probability of remaining in the atomic state in the limit $t\to +\infty$ is 
\begin{equation}
	P_a = \exp\left( -2\pi g^2 / \alpha \right).
\end{equation}

\begin{table}[h!]
	\begin{center}
		\caption{Overview of the calculated and fitted coupling coefficients and the resulting enhancement factor for the respective temperature. }
		\label{tab:coupling2}
		\scriptsize
		\begin{tabular}{@{}cccccccc} 
			\br
			\text{Fermi Temp.} & \text{Temp.} & \text{Calculated $g$} & \text{Fitted $g$} & Enhancement & Peak density & Mode volume & Adiabatic rate \\
			$T_F~({\rm \mu K})$ & $T/T_F$ & $g_c~(2\pi\hbar\times{\rm kHz})$ & $g_f~(2\pi\hbar\times{\rm kHz})$ & $g_f^2/g_c^2$ & $\rho~(10^{12}{\rm cm}^{-3})$ & ${\cal V}$ $(10^4{\rm \mu m}^3)$ & $\Gamma_{\infty}$ \\
			\mr
			1.23 (1) & 0.11 (1) & 4.79 & 8.78 (42) & 3.4  (6)  & 6.28 & 5.75 & 0.77 (2) \\
			1.30 (1) & 0.20 (1) & 4.62 & 7.68 (14) & 2.76 (16) & 7.21 & 6.18 & 0.74 (2) \\
			1.39 (1) & 0.43 (2) & 4.48 & 6.46 (22) & 2.08 (20) & 6.35 & 6.59 & 0.73 (2) \\
			1.48 (1) & 0.50 (2) & 4.30 & 5.31 (29) & 1.52 (21) & 6.50 & 7.14 & 0.74 (5) \\
			1.53 (3) & 0.62 (1) & 4.30 & 4.51 (42) & 1.10 (21) & 5.29 & 7.14 & 0.76 (1) \\
			1.61 (3) & 0.86 (2) & 4.19 & 4.55 (27) & 1.18 (15) & 5.39 & 7.54 & 0.63 (4) \\
			1.69 (2) & 1.23 (7) & 4.69 & 5.95 (29) & 1.61 (20) & 5.11 & 6.00 & 0.49 (2) \\
			1.86 (3) & 1.73 (4) & 4.73 & 4.64 (47) & 0.96 (19) & 5.39 & 5.91 & 0.37 (3) \\
			\br
		\end{tabular}
	\end{center}
\end{table}

To obtain the volume, we note that
typically two-body interactions will change the shape and density distribution of atoms in the trap. Papenbrock and Bertsch~\cite{Papenbrock1999} introduced a parameter $\xi_B$ such that the chemical potential is scaled by the Fermi energy of the non-interacting case $\mu = \xi_B E_F^0$. The trapping frequency is then scaled by $\sqrt{\xi_B}\omega_i$ accounting for the change of effective trapping frequency. Then the radii of the atomic cloud read
\begin{equation}
	R_i = \xi_B^{1/4} a_{ho} \frac{\omega_{ho}}{\omega_i} (24N)^{1/6},
\end{equation}
yielding the volume of a spherical gas,
\begin{equation}
	{\cal V} = \frac{4\pi}{3} a_{ho}^3 \xi_B^{3/4} \sqrt{24N}.
\end{equation}
In the BCS regime, $\xi_B$ is calculated by the Monte Carlo method~\cite{Carlson2011}. Though depending on trapping profile and particle density, $\xi_B$ converges to $\approx 0.37$ in dilute limit, which is used in the calculation (table~\ref{tab:coupling2}).

\section{Many-body model of the atom-molecule coupling}

The formation of bosonic molecules from pairs of fermionic atoms is modeled by a spin-boson coupled system~\cite{javanainen_collective_2004-1,pazy_nonlinear_2005,pazy_conversion_2004}. The total Hamiltonian consists of different molecular states, and for a particular molecular state annihilated by $\hat{b}$, it reads
\begin{equation}
	\label{aeq:Original_Hamiltonian}
	H = \delta(t) \hat{b}^\dag \hat{b} + \sum_j \varepsilon_{j} \left( \hat{c}_{j\uparrow}^\dag\hat{c}_{j\uparrow} + \hat{c}_{j\downarrow}^\dag\hat{c}_{j\downarrow} \right)
	+ g \left( \hat{b}^\dag \sum_j \hat{c}_{j\downarrow}\hat{c}_{j\uparrow} + \text{H.c.} \right). \nonumber
\end{equation}
Here $\delta(t)=\Delta\mu(\alpha t  + B_0)$ gives the molecular energy, where $\alpha$ and $B_0$ are the ramping rate and the initial value of the magnetic field. The paramter $\varepsilon_{j}$ denotes the kinetic and trap energy of the atom pair. Typically, it can be the harmonic levels $\varepsilon_{j}=\hbar\omega(j+1/2)$, or free space by replacing $j$ by $\mathbf{k}$, $\varepsilon_\mathbf{k}=\hbar^2 k^2/2m$. The molecular energy is given by $\delta(t)$. 

When sweeping the magnetic field from $860~{\rm G}$ to $707~{\rm G}$, the molecular energy changes from $\delta(t_0)/\hbar=-2\pi\times72.7~{\rm MHz}$ to $\delta(t_f)/\hbar=2\pi\times355~{\rm MHz}$. The maximum value of $\varepsilon_j$ (Fermi level) is roughly $\varepsilon_F = \hbar^2/(2m) (3\pi^2n)^{2/3} \approx 2\pi\hbar \times 8.03~{\rm kHz}$ for a density of $n=10^{12}~{\rm cm}^{-3}$. Taking the full range of the magnetic field, the numerical cost in the simulation will be very expensive. To simplify the calculation, we have chosen an initial value of magnetic field relatively close to the resonance, which captures the LZ transition dynamics.

\begin{figure} [t]
	\centering
	\includegraphics[width=0.5\linewidth]{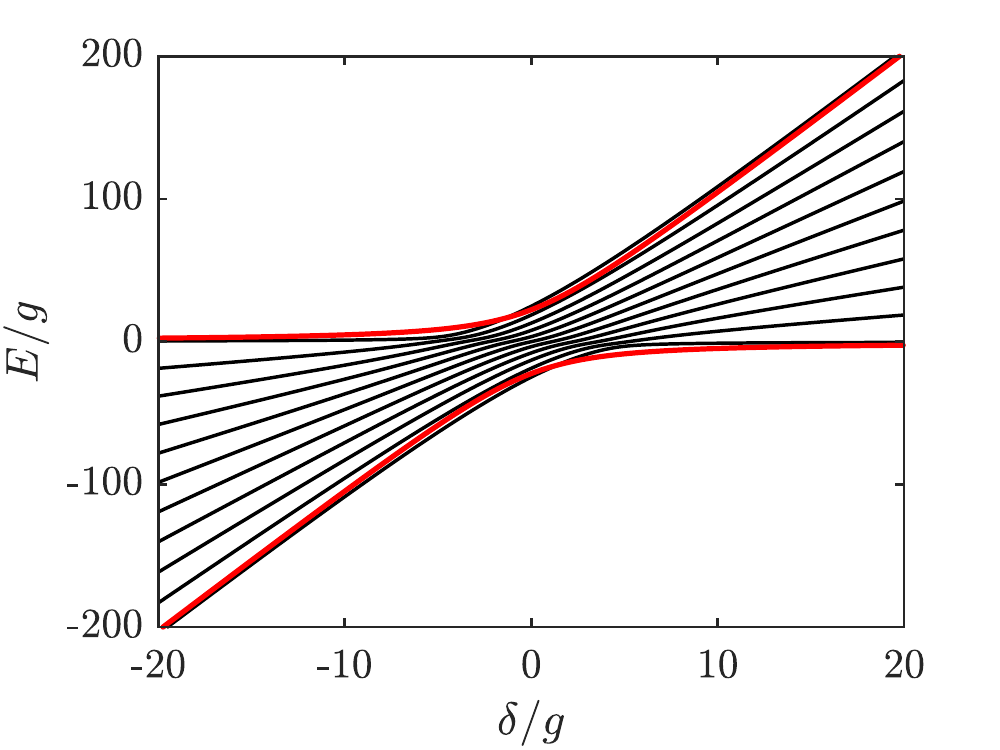}
	\caption{\footnotesize(Color online) \textbf{Instantaneous energy spectrum.} (black) The eigenstate energies for Dicke state basis \eqref{eq:Dicke} with $N=10$. (red) The energy spectrum for the effective Hamiltonian \eqref{eq:effective}. }
	\label{fig:S2}
\end{figure}

As we ignore dephasing and disassociation of the molecular states, the system is closed. We will use the Dicke state basis due to the $U(1)$ symmetry (i.e. particle number conservation), which will significantly reduce the Hilbert space. For $2N$ atoms, there are $N+1$ states encoded by the quantum number $m$ (equal to the number of molecules), such that
\begin{equation}
	\left| N; m \right\rangle = 1/\sqrt{C_N^m} \sum_{ x \in \{0,1\}^N \ \& \ {\rm wt}(x)=m } \left| m, x \right\rangle,
\end{equation}
where $x$ refers to $N$-qubit state, $C_N^m$ is the combination number, and ${\rm wt}(x)$ is the Hamming weight (the number of non-zero bits in the sequence $x$). For example, $|3;2\rangle = 1/\sqrt{3} \left( |2,110\rangle + |2,101\rangle + |2,011\rangle \right)$. Then we show that the Dick state basis is closed for the Hamiltonian equation~\eqref{aeq:Original_Hamiltonian}, and we only need to focus on the off-diagonal part, 
\begin{align*}
	\left\langle N, m+1 \middle| \hat{b}^\dag \sum_j \hat{c}_{j\downarrow} \hat{c}_{j\downarrow} \middle| N, m \right\rangle &= \frac{1}{\sqrt{C_N^{m+1} C_N^m}} \sqrt{m+1} C_N^{m} \\
	&= (m+1) \sqrt{N-m}.
\end{align*}

The matrix form of the Hamiltonian reads
\begin{equation} \label{eq:Dicke}
	H = \begin{pmatrix}
		0			&	g\sqrt{N}		&	0				&	\cdots	&	0				&	0			\\
		g\sqrt{N}	&	\delta(t)		&	g2\sqrt{N-1}	&	\cdots	&	0				&	0			\\
		0			&	g2\sqrt{N-1}	&	2\delta(t)		&	\cdots	&	0				&	0			\\
		\vdots		&	\vdots			&	\vdots			&	\ddots	&	\vdots			&	0			\\
		0			&	0				&	0				&	\cdots	&	(N-1) \delta(t)	&	gN			\\
		0			&	0				&	0				&	\cdots	&	gN				&	N \delta(t)			
	\end{pmatrix}.
\end{equation}
The instantaneous energy spectrum is shown in figure~\ref{fig:S2}.

A comparison of the two-level model with the Hamiltonian 
\begin{equation} \label{eq:effective}
	H = \begin{pmatrix}
		0			&	g N^{3/2}	\\
		g N^{3/2}	&	N \delta(t)	\\
	\end{pmatrix}
\end{equation}
are shown. The spectra of the two-level model agree with the upper and lower bound of the Dick states, and the energy gap is scaled by $\sim N^{3/2}$, and the time is scaled by $\sim N^{-1}$. 

\section*{References}

%

\end{document}